\documentclass[onecolumn,a4paper]{article}
\usepackage{syntonly}
\usepackage[tbtags]{amsmath}\numberwithin{equation}{section}
\usepackage{amssymb}
\usepackage{curves}
\usepackage{float}

\begin{document}
\title{The Lie Group Structure of the $\eta-\xi$ Space-time and its
Physical Significance}
\author{Gu Zhiming\\College of Science, Nanjing University of
Aeronautics and Astronautics\\Nanjing 210016,China}
\date{March 1,2004}
\maketitle
\begin{abstract}
    The $\eta-\xi$ space-time is suggested by Gui for the
quantum field theory in 1988. This paper consists of two parts.
The first part is devoted to the discussion of the global
properties of the $\eta-\xi$ space-time. The result contains a
proof which asserts that the $\eta-\xi$ space-time is homeomorphic
to $\mathbb{C}^{*}\times \mathbb{C}^{*}\times \mathbb{C}^{2}$ by
means of two explicit maps, which shows that the $\eta-\xi$
space-time allows a Lie group structure. Thus some transformation
groups, one of which is isomorphic to the Lorentz group in two
dimensions, can be found. The other part of the paper is the
discussion about the embedding of some subspaces in the $\eta-\xi$
space-time. In particular, it is pointed out that the Euclidean
space-time and the Minkowskian space-time are linked in a way in
the $\eta-\xi$ space-time such that the tilde field appears
naturally. In addition some formulae in the $\eta-\xi$ space-time
reappear in a more natural
way.
\end{abstract}

\section{Introduction}
    \indent
     The $\eta-\xi$ space-time(Gui's space-time)is proposed by Gui for
the quantum field theory in [1],[2] and [3].In his resent paper
[4] the idea and progress on this new space-time are reviewed and
some problems to be solved are given as well. There are two
problems among that encouraged by [4], the symmetry problem and
tilde field problem, in which we are interested.\\
    \indent In the paper the fact that the $\eta-\xi$ space-time is homeomorphic
to $\mathbb{C}^{*}\times \mathbb{C}^{*}\times \mathbb{C}^{2}$ by
means of two explicit maps, one of which gives the $\eta-\xi$
space-time the Lie group structure directly, is showed. From these
results some excellent formulae in the theory of $\eta-\xi$
space-time reappear in a more natural way. Next it comes out that
the Euclidean space-time $\mathbb{S}^1\times\mathbb{R}^3$ for the
imaginary-time theory and the Minkowskian space-time
$\mathbb{R}^4$ for the real-time theory join as two subspaces of
the $\eta-\xi$ space-time. This property leads the tilde field to
appear naturally.\\
    \indent The $\eta-\xi$ space-time is denoted by $V$ which is
defined in our way by
               $V=\mathbb{C}^4-C$
where an element in $\mathbb{C}^4$ is denoted by four complex
numbers $(\eta,\xi,y,z)$, and $C$ is the algebraic set determined
by the equation\\
    \begin{equation*}
    \xi^2-\eta^2=0
    \end{equation*}
with the metric $g$ which is defined by
    \begin{equation*}
    ds^{2}=\frac{1}{\alpha^{2}(\xi^{2}-\eta^{2})}(-d\eta^{2}+d\xi^{2})+dy^{2}+dz^{2}
    \end{equation*}
where $\alpha\neq0$, is a real constant number. There is no loss
in generality by assuming that $\alpha=1$ in this paper.\\

    \indent
    In order to define the metric $g$ on the whole space-time we
delete the algebraic set $C$ from $\mathbb{C}^4$.\\

    \indent
    Let $V_{0}=\mathbb{C}^2-C_{0}$,where $C_{0}$ is the algebraic
set determined by the equation\\
    \begin{equation*}
    \xi^2-\eta^2=0
    \end{equation*}
with the metric $g_{0}$ which is defined by
    \begin{equation*}
    ds^{2}=\frac{1}{\xi^{2}-\eta^{2}}(-d\eta^{2}+d\xi^{2})
    \end{equation*}
It is obvious that $V=V_{0}\times\mathbb{C}^2$. thus it suffices
to research the topology of $V_{0}$.

   \indent
    We have need of a little knowledge of Lie groups,for these see
[5] or [6].

\section{ The structure of $V_{0}$}
\indent
   In order to give $V_{0}$ a group structure we consider the
subgroup $T$ of $GL(2,\mathbb{C})$, that is\\
   \begin{equation*}
     T=\{\begin{pmatrix}\gamma&0\\0&\delta\end{pmatrix}|\gamma\delta\neq0\}
     \end{equation*}
It is obviously homeomorphic to
$\mathbb{C}^{\ast}\times\mathbb{C}^{\ast}$. Then we have a map \\
    \begin{equation*}V_{0}\rightarrow T\end{equation*}
    \begin{equation*}(\eta,\xi)\mapsto\begin{pmatrix}\xi-\eta&0\\0&\xi+\eta\end{pmatrix}\end{equation*}
It is a homeomorphism map. If identify $(\eta,\xi)$ in $V_{0}$
with the element \\
         \begin{equation*}\begin{pmatrix}\xi&\eta\\\eta&\xi\end{pmatrix}\in GL(2,\mathbb{C})\end{equation*}
then $V_{0}$ can be considered as a subgroup, which is a conjugate
of $T$,of $GL(2,\mathbb{C})$. In fact we have \\
\begin{equation*}
  \begin{pmatrix}\frac{1}{\sqrt{2}}&\frac{1}{\sqrt{2}}\\-\frac{1}{\sqrt{2}}&\frac{1}{\sqrt{2}}\end{pmatrix}^{-1}
  \begin{pmatrix}\xi&\eta\\\eta&\xi\end{pmatrix}\begin{pmatrix}\frac{1}{\sqrt{2}}&\frac{1}{\sqrt{2}}\\
  -\frac{1}{\sqrt{2}}&\frac{1}{\sqrt{2}}\end{pmatrix}=\begin{pmatrix}\xi-\eta&0\\0&\xi+\eta\end{pmatrix}.
\end{equation*}
  \indent
   Having the above identification, we define a multiplication in
$V_{0}$ as follows\\
       \begin{equation*}(\eta,\xi)\ast(\eta\prime,\xi\prime)=(\xi\eta\prime+\eta\xi\prime,\xi\xi\prime+\eta\eta\prime)\end{equation*}
It is easily checked that $(0,1)$ is the identity, the inverse
element of $(\eta,\xi)$ is
$(\frac{-\eta}{\xi^{2}-\eta^{2}},\frac{\xi}{\xi^{2}-\eta^{2}})$
and the multiplication is associative and commutative. In
addition, there is a representation\\
  \begin{equation*}\Phi:V_{0}\rightarrow GL(2,\mathbb{C})\end{equation*}
  \begin{equation*}\Phi(\eta,\xi)=\begin{pmatrix}\xi&\eta\\\eta&\xi\end{pmatrix}\end{equation*}
In fact, it is the above identification. In addition we have
also\\
 \begin{equation*}(\eta,\xi)\ast(\eta\prime,\xi\prime)=(\eta,\xi)\begin{pmatrix}\xi\prime&\eta\prime\\\eta\prime&\xi\prime\end{pmatrix}\end{equation*}
 \indent
As a group ,$V_{0}$ has the left translation, that is\\
\centerline{$L_{(\gamma,\delta)}:V_{0}\rightarrow V_{0}$}
   \begin{equation*}L_{(\gamma,\delta)}(\eta,\xi)=(\gamma,\delta)\ast(\eta,\xi)\end{equation*}\\
for a fixed element $(\gamma,\delta)\in V_{0}$. It is easy to show
that the left translations preserve the metric $g_{0}$. It follows
that the set of the left translations  forms a transformation
group , which  is isomorphic  to $V_{0}$,of $V_{0}$.\\
\indent Let\\
$$G_{1}=\{(\gamma,\delta)\in V_{0}|\delta^{2}-\gamma^{2}=1\}\quad\text{and}\quad G_{2}=\{(\gamma,\delta)\in G_{1}|\gamma,\delta\in\mathbb{R}\}$$\\
it is easy to show that $G_{1}$  is a subgroup of the group
$V_{0}$ and $G_{2}$ is a subgroup of $G_{1}$ . Since $\Phi(G_{2})$
is the Lorentz group in two dimensions, $\eta-\xi$ space-time has
a transformation group which is isomorphic to the Lorentz group in
two dimensions.\\
   \indent
    As a commutative Lie group,$V_{0}$  has its Lie algebra
$\mathcal{V}_{0}$ with  the trivial  bracket , which is isomorphic
to $\mathbb{C}^{2}$ by taking a basis, for example
$$\frac{\partial}{\partial\eta},\frac{\partial}{\partial\xi}$$  This
fact allows us to define the exponential  map\\
\centerline{$\exp:\mathbb{C}^{2}\rightarrow V_{0}$}
\begin{equation*}\begin{split}
    \exp(z_{0},z_{1})&=\Phi^{-1}\exp\begin{pmatrix}z_{1}&z_{0}\\z_{0}&z_{1}\end{pmatrix}\\
                     &=\Phi^{-1}(\exp\begin{pmatrix}z_{1}&0\\0&z_{1}\end{pmatrix}\exp\begin{pmatrix}0&z_{0}\\z_{0}&0\end{pmatrix})\\
                     &=\Phi^{-1}\begin{pmatrix}e^{z_{1}}\cosh z_{0}&e^{z_{1}}\sinh z_{0}\\e^{z_{1}}\sinh z_{0}&e^{z_{1}}\cosh z_{0}
                     \end{pmatrix}\\
                     &=(e^{z_{1}}\sinh z_{0},e^{z_{1}}\cosh z_{0})
\end{split}
\end{equation*}
This expression and that of [1] are coincide exactly. Then we find
out \\
\begin{equation}
   \frac{1}{\xi^{2}-\eta^{2}}(-d\eta^{2}+d\xi^{2})=-dz_{0}^{2}+dz_{1}^{2}
\end{equation}
\indent
    Since $V_{0}$  is a commutative Lie group,the map $\exp $ is a  homomorphism  from
the additive group of $\mathbb{C}^{2}$ to $V_{0}$  with the
composition $\ast$. Thus the translation
\textquoteleft\textquoteleft$(a,b)+$\textquoteright\textquoteright
in $\mathbb{C}^{2}$ corresponds to the translation
\textquoteleft\textquoteleft$(\gamma,\delta)\ast$\textquoteright\textquoteright
in $V_{0}$, where $(\gamma,\delta)=\exp(a,b)$.
\indent
    It follows from the connectedness of $V_{0}$ that $\tilde{Q}=\exp:\mathbb{C}^{2}\rightarrow V_{0}$
is the  universal  covering  map of $V_{0}$. Thus there is another
homeomorphism, if write
$\mathbb{C}^{2}=\mathbb{R}_{0}\times\mathbf{i}\mathbb{R}_{0}\times\mathbb{R}_{1}\times\mathbf{i}\mathbb{R}_{1}$,
\begin{equation*}Q:(\mathbb{R}_{0}\times\mathbb{S}^{1}_{0})\times(\mathbb{R}_{1}\times\mathbb{S}^{1}_{1})\rightarrow
V_{0}\end{equation*}
\begin{align}
Q(u_{0},e^{\mathbf{i}v_{0}},u_{1},e^{\mathbf{i}v_{1}})=\tilde{Q}(u_{0}+\mathbf{i}v_{0},u_{1}+\mathbf{i}v_{1})
&\quad\quad& 0\leq v_{0},v_{1}<2\pi
\end{align}
and a map $\Pi:\mathbb{C}^{2}\rightarrow
(\mathbb{R}_{0}\times\mathbb{S}^{1}_{0})\times(\mathbb{R}_{1}\times\mathbb{S}^{1}_{1})$
such that $\tilde{Q}=Q\circ\Pi$ ,the composition of two maps.\\

\section{The embedding of subspaces}
\indent
 Restricting $Q$  to $\{t\}\times\mathbb{S}^{1}_{0}\times\mathbb{R}_{1}\times\{1\}$ for a fixed $t$, we  have an
 embedding\\
 \begin{equation*}Q_{I,t}:\mathbb{S}^{1}_{0}\times\mathbb{R}_{1}\rightarrow
 V_{0}\end{equation*}
 \begin{align}
Q_{I,t}(e^{\mathbf{i}\tau},x_{1})&=Q(t,e^{\mathbf{i}\tau},x_{1},1)\notag\\
                                  &=(e^{x_{1}}\sinh(t+\mathbf{i}\tau),e^{x_{1}}\cosh(t+\mathbf{i}\tau))
\end{align}
In particular, if $t=0$,$
Q_{I,0}(e^{\mathbf{i}\tau},x_{1})=(\mathbf{i}e^{x_{1}}\sin\tau,e^{x_{1}}\cos\tau)$.\\
\indent If $\tau$ is fixed but $t$ is variable we have another
embedding\\
\begin{equation*}Q_{R,\tau}:\mathbb{R}_{0}\times\{e^{\mathbf{i}\tau}\}\times\mathbb{R}_{1}\rightarrow
V_{0}\end{equation*}
\begin{align}
 Q_{R,\tau}(t,x_{1})&=\tilde{Q}(t+\mathbf{i}\tau,x_{1})
                    &=(e^{x_{1}}\sinh (t+\mathbf{i}\tau),e^{x_{1}}\cosh (t+\mathbf{i}\tau))
\end{align}
Let $(\sinh\mathbf{i}\tau',\cosh\mathbf{i}\tau')\in V_{0}$, then \\
\begin{multline*}
(\sinh\mathbf{i}\tau',\cosh\mathbf{i}\tau')\ast(e^{x_{1}}\sinh(t+\mathbf{i}\tau),e^{x_{1}}\cosh(t+\mathbf{i}\tau))
\\=(e^{x_{1}}\sinh (t+\mathbf{i}(\tau+\tau')),e^{x_{1}}\cosh
(t+\mathbf{i}(\tau+\tau')))
\end{multline*}
It follows that
$Q_{R,\tau+\tau'}(\mathbb{R}_{0}\times\{e^{\mathbf{i}(\tau+\tau')}\}\times\mathbb{R}_{1})$
can  be  obtained    from
$Q_{R,\tau}(\mathbb{R}_{0}\times\{e^{\mathbf{i}\tau}\}\times\mathbb{R}_{1})$
by translation along
$(\sinh\mathbf{i}\tau',\cosh\mathbf{i}\tau')$. In Particular,
after the translation along $(0,-1)$ , we have\\
 \begin{align*}
 Q_{R,\tau+\pi}(t,x_{1})&=(-e^{x_{1}}\sinh (t+\mathbf{i}\tau),-e^{x_{1}}\cosh
 (t+\mathbf{i}\tau))\\
                        &=-Q_{R,\tau}(t,x_{1})
\end{align*}
For example,when $\tau=0$,we have\\
\begin{align*}
Q_{R,0}(t,x_{1})&=(e^{x_{1}}\sinh t,e^{x_{1}}\cosh t)\\
Q_{R,\pi}(t,x_{1})&=(-e^{x_{1}}\sinh t,-e^{x_{1}}\cosh t)
\end{align*}
The subspaces
$Q_{R,0}(\mathbb{R}_{0}\times\{1\}\times\mathbb{R}_{1})=V_{I}$ and
$Q_{R,\pi}(\mathbb{R}_{0}\times\{-1\}\times\mathbb{R}_{1})=V_{II}$,as
is known, are called  the  universe $I$ and the universe $II$.\\
\indent
    Next  the  line  elements  are  calculated. From  (2.1),(2.2) and (3.1),
we  have\\
\begin{equation}
ds^{2}=d\tau^{2}+dx_{1}^{2}
\end{equation}
This  is  the  Euclidean  line  element  in
$Q_{I,t}(\mathbb{S}^{1}_{0}\times\mathbb{R}_{1})$.From (2.1) and
(3.2) we have \\
\begin{equation}
ds^{2}=-dt^{2}+dx_{1}^{2}
\end{equation}
This  is  the  Minkowskian  line  element  in
$Q_{R,\tau}(\mathbb{R}_{0}\times\{e^{\mathbf{i}\tau}\}\times\mathbb{R}_{1})$.\\
\indent
Let $v=(a,b)\in\mathbb{C}^{2}$,the Lie algebra of
$V_{0}$,we have the one-parameter subgroup\\
\begin{equation*}
\rho_{v}(\mu)=\exp\mu v=(e^{\mu b}\sinh\mu a,e^{\mu b}\cosh\mu a).
\end{equation*}
It induces  a  one-parameter transformation  group\\
\begin{equation*}
\rho_{v}(\mu)\ast(\eta,\xi)=e^{\mu b}(\eta\cosh\mu a+\xi\sinh\mu
a,\xi\cosh\mu a+\eta\sinh\mu a)
\end{equation*}
and its generator\\
\begin{equation}
(\eta,\xi)\mapsto(b\eta+a\xi,b\xi+a\eta)
\end{equation}
Because  of  metric- preserving  the  generator  is  a  Killing
field.\\
\indent As the tangent vector of $V_{0}$ , a vector $v$  in
$\mathcal{V}_{0}$  has the same causal character as that of the
Killing field induced by $v$.In particular, if $v$ is the
time-like  vector $\frac{\partial}{\partial\eta}$, see
Equ.(2.1), then we have a time-like Killing field, see Equ.(3.5),\\
\centerline{$\xi\frac{\partial}{\partial\eta}+\eta\frac{\partial}{\partial\xi}$}
which is exactly what [2] wanted.\\

   In  conclusion  we  point  out  that\\
 (1)\ \ $\mathbb{C}^{2}$ is  the  universal covering space  of
$V_{0}$.\\
 (2)\ \ $Q_{I,t}(\mathbb{S}^{1}_{0}\times\mathbb{R}_{1})$
 have the Euclidean  line element and the universal
covering space  $\mathbf{i}\mathbb{R}_{0}\times\mathbb{R}_{1}$ for
every $t$.\\
 (3)\ \ $Q_{R,\tau}(\mathbb{R}_{0}\times\{e^{\mathbf{i}\tau}\}\times\mathbb{R}_{1})$
 have the Minkowskian line element, in particular,$Q_{R,0}(\mathbb{R}_{0}\times\{1\}\times\mathbb{R}_{1})=V_{I}$
 and $Q_{R,\pi}(\mathbb{R}_{0}\times\{-1\}\times\mathbb{R}_{1})=V_{II}$
 are the universe $I$ and the universe $II$,respectively.

\section{The imaginary-time and the real-time}
    Let $\phi$ is a field  on  $V_{0}$. The imaginary-time theory
considers\\
$$\phi_{0}:\mathbb{S}^{1}_{0}\times\mathbb{R}_{1}\xrightarrow[]{Q_{I,t}}V_{0}\xrightarrow[]{\phi}\mathbb{R}$$\\
and  the  real-time  theory  considers\\
$$\phi_{1}:\mathbb{R}_{0}\times\{1\}\times\mathbb{R}_{1}\xrightarrow[]{Q_{R,0}}V_{0}\xrightarrow[]{\phi}\mathbb{R}$$\\
$$\phi_{2}:\mathbb{R}_{0}\times\{-1\}\times\mathbb{R}_{1}\xrightarrow[]{Q_{R,\pi}}V_{0}\xrightarrow[]{\phi}\mathbb{R}$$\\

\restylefloat{figure}
\begin{figure}[H]
\setlength{\unitlength}{1cm}
\begin{picture}(9,8)
\put(1,6){\vector(1,0){6.5}} \put(4.5,1){\vector(0,1){6.5}}
\linethickness{2pt}
\put(2,6){\line(1,0){5}}\put(2,3.5){\line(1,0){5}}\put(7,3.5){\line(0,1){2.5}}\put(2,1){\line(0,1){2.5}}
\put(2,6.5){\makebox(0,0)[l]{$-F$}}\put(7,6.5){\makebox(0,0)[r]{$F$}}\put(2,4){\makebox(0,0)[r]{$-F-\frac{\mathbf{i}\beta}{2}$}}\put(2,1.5){\makebox(0,0)[r]{$-F-\mathbf{i}\beta$}}
\put(7,3){\makebox(0,0)[l]{$F-\frac{\mathbf{i}\beta}{2}$}}\put(4.2,7.5){\makebox(0,0)[r]{$Im$}}\put(8,6){\makebox(0,0)[l]{$Re$}}
\put(5,5.8){\makebox(0,0)[t]{$C_{1}$}}\put(7,4.75){\makebox(0,0)[l]{$C_{3}$}}\put(5,3.3){\makebox(0,0)[t]{$C_{2}$}}\put(2,2.5){\makebox(0,0)[l]{$C_{4}$}}
\end{picture}
\caption{The time path $C'$ in
$\mathbb{C}=\mathbb{R}_{0}\times\mathbf{i}\mathbb{R}_{0}$}
\end{figure}
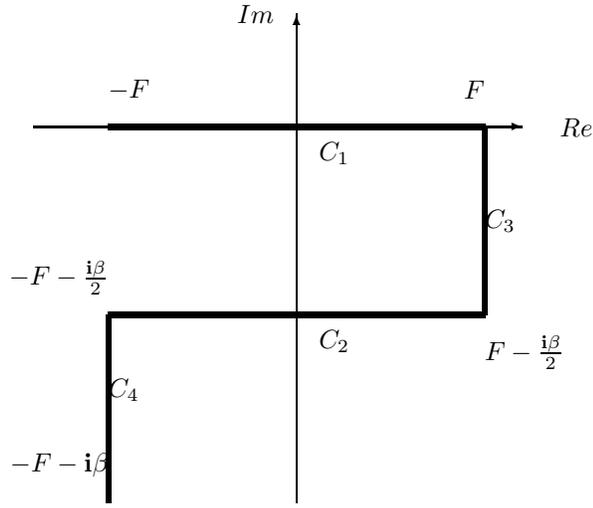

\restylefloat{figure}
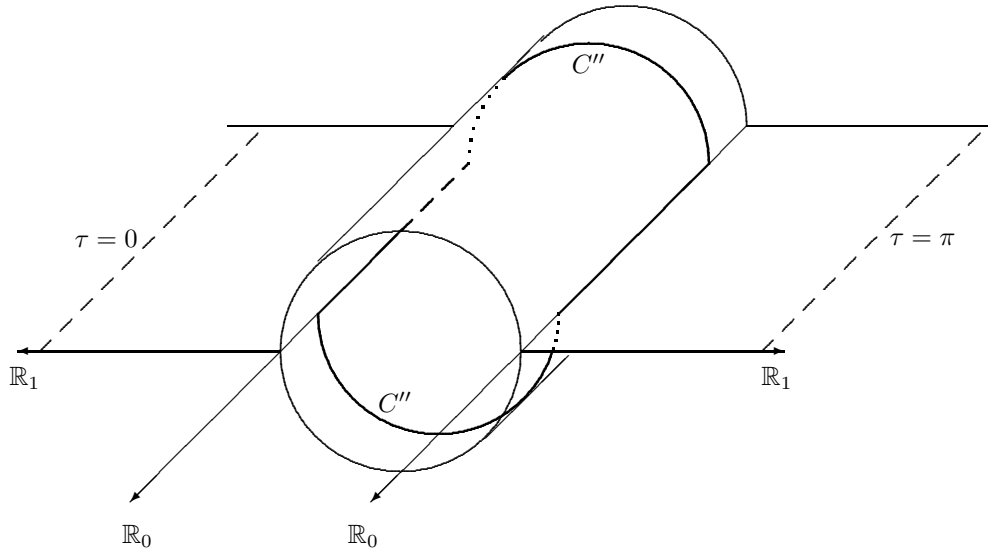
\begin{figure}[H]
\setlength{\unitlength}{1cm}
\begin{picture}(14,8)
\put(5.5,3){\bigcircle{3.2}}\put(8.5,6){\arc(1.6,0){135}}
\put(10.1,6){\vector(-1,-1){5}}\put(4.4,3.5){\vector(-1,-1){2.5}}\put(7.1,3){\vector(1,0){3.5}}\put(3.9,3){\vector(-1,0){3.5}}
\put(10.1,6){\line(1,0){3.2}}\put(6.2,6){\line(-1,0){3}}\put(4.4,4.2){\line(1,1){3}}\put(6.63,1.85){\line(1,1){1.1}}
\thicklines
\put(8,5.5){\arc(1.6,0){135}}\put(8,5.5){\arc[4](-1.13,1.13){45}}\put(6,3.5){\arc(-1.6,0){163}}\put(6,3.5){\arc[4](1.53,-0.47){17}}
\put(7.6,3.5){\line(1,1){2}}\put(4.4,3.5){\line(1,1){1.13}}
\curvedashes{0,0.2,0.2}\curve(5.5,4.6,6.4,5.5)
\put(8,6.7){\makebox(0,0)[b]{$C''$}}\put(5.2,2.2){$C''$}\put(12,4.5){\makebox(0,0)[l]{$\tau=\pi$}}\put(2,4.5){\makebox(0,0)[r]{$\tau=0$}}\put(10.5,2.8){\makebox(0,0)[t]{$\mathbb{R}_{1}$}}
\put(0.5,2.8){\makebox(0,0)[t]{$\mathbb{R}_{1}$}}\put(5,0.7){\makebox(0,0)[t]{$\mathbb{R}_{0}$}}\put(2,0.7){\makebox(0,0)[t]{$\mathbb{R}_{0}$}}
\thinlines
\curvedashes{0,0.2,0.2}\curve(10.3,3,13.3,6)\curvedashes{0,0.2,0.2}\curve(0.7,3,3.7,6)
\end{picture}
\caption{$C''=\Pi(C')$ in
$(\mathbb{S}^{1}_{0}\times\mathbb{R}_{1})\cup\bigcup_{\tau=0,\pi}(\mathbb{R}_{0}\times\{e^{\mathbf{i}\tau}\}\times\mathbb{R}_{1})$}
\end{figure}
    In order  to  link  the imaginary-time  and the real-time note
that $Q_{I,t}(\mathbb{S}^{1}_{0}\times\mathbb{R}_{1})$  share the
space $\mathbb{R}_{1}$  with
$Q_{R,\tau}(\mathbb{R}_{0}\times\{e^{\mathbf{i}\tau}\}\times\mathbb{R}_{1})$,
$\tau=0,\pi$, in$V_{0}$. Now we look at the calculation of  the
real-time thermal Green's  functions  by computing  a path
integral along the time path $C'$ in $\mathbb{C}$,where
$C'$consists of $C_{1}$,$C_{2}$,$C_{3}$ and $C_{4}$ ,( Figure  1 ,
see[ 7 ]).
     We can identify the "time-plane" $\mathbb{C}$ with
$\mathbb{R}_{0}\times\mathbf{i}\mathbb{R}_{0}$ , then let
$C''=\Pi(C')$, see Figure 2. It is easy to find\\
\indent (1)  If $F\rightarrow 0$, then $C''$ will be deformable in\\
\centerline{$(\mathbb{S}^{1}_{0}\times\mathbb{R}_{1})\cup\bigcup_{\tau=0,\pi}(\mathbb{R}_{0}\times\{e^{\mathbf{i}\tau}\}\times\mathbb{R}_{1})$}
into a circle and the imaginary-time theory  is obtained.\\
\indent (2)  If $F\rightarrow +\infty$, then $\phi_{1}$ and
$\phi_{2}$
are obtained, where $\phi_{2}$ is the tilde field.\\

\section{An open problem}
\indent Because $\eta-\xi$ space-time is homotopy equivalent to
the torus, it has nontrivial cohomology groups,
precisely,$H^{1}(V_{0};\mathbb{Z}_{2})=\mathbb{Z}_{2}\oplus\mathbb{Z}_{2}$,
$H^{2}(V_{0};\mathbb{Z})=\mathbb{Z}$ . As a result of these
properties ,there are nontrivial  real and complex line bundles
over$V_{0}$. It seems to me that the physical significance of
these facts is unknown.\\

\end{document}